\documentstyle[amstex,aps,prb,multicol,epsfig,amssymb,eqsecnum,picinpar]{revtex}
\begin{document}
\draft
\title{The effects of interactions and disorder in the two-dimensional 
chiral metal}
\author{J. J. Betouras and J. T. Chalker}
\address{
Theoretical Physics,
Oxford University,
1 Keble Road,
Oxford, OX1 3NP,
United Kingdom}
\date{\today}
\maketitle

\begin{abstract}
We study the two-dimensional chiral metal, 
which is formed at the surface of a layered
three-dimensional system exhibiting the integer quantum Hall
effect by hybridization of the edge states associated with each layer
of the sample. We investigate mesoscopic fluctuations, dynamical screening 
and inelastic scattering in the chiral metal, focussing particularly on 
fluctuations of conductance, $\delta g(B)$,
with magnetic field, $B$. The correlation function 
$\langle \delta g(B) \delta g(B+\delta B)\rangle$ provides information
on the inelastic scattering rate, $\tau_{in}^{-1}$, through both the variance
of fluctuations and the range of correlations in $\delta B$. We calculate 
this correlation function for samples which are not fully phase coherent. 
Two
regimes of behaviour exist, according to whether $\tau_{in}^{-1}$ is
smaller or larger than $\tau_{\perp}^{-1}$, the rate for inter-edge
tunneling, and we give results in both regimes.
We also investigate dynamical screening of Coulomb interactions in the 
chiral metal and calculate the contribution to $\tau_{in}^{-1}$ from 
electron-electron scattering, finding $\tau_{in}^{-1} \propto T^{3/2}$ 
for $\tau_{in}^{-1} \ll \tau_{\perp}^{-1}$ at temperature $T$.
\end{abstract}
\pacs{PACS numbers: 73.40.Hm, 73.23.-b, 73.20.-r, 72.15.Lh}

\begin{multicols}{2}

\section{Introduction}

An unusual type of two-dimensional conductor, the chiral metal,
may be formed at the surface of a layered, three-dimensional system in the 
presence of a magnetic field which has a component perpendicular to the
layers \cite{john,balents,druist}. 
It occurs in a system 
of weakly-coupled layers at magnetic 
field-strengths for which an isolated layer would have an integer
quantised Hall conductance. Under these circumstances, bulk states
are localised and the extended, chiral edge states associated
with each layer hybridise to form a two-dimensional conductor
at the surface of the sample. Properties of this chiral metal may be probed
by conductance measurements with current flow normal to the layers.

Early experimental work on semiconductor multilayer samples \cite{stormer} 
established that the integer quantum Hall effect of a single 
layer is indeed robust against weak interlayer tunneling, 
and suggested that, within a quantum Hall plateau, interlayer
conductivity in the bulk vanishes in the low-temperature limit.  
Subsequent theoretical discussions \cite{john,balents} drew attention to
the unusual nature of surface states in multilayer quantum 
Hall systems, and some aspects of the theory of the chiral metal have 
since been explored in considerable detail 
\cite{kim,mathur,yu,bfz,grs1,grs2,cho,wang,wangplerou,meir,johnsondhi,betouras}. 
Recent experiments by 
Druist {\it et al} have isolated
the contribution of surface states to vertical transport \cite{druist}
and demonstrated that mesoscopic conductance fluctuations offer a way 
to study inelastic scattering in the chiral metal \cite{druist2,gwinntalk}. 
Conduction by surface states 
has
also been invoked in the interpretation of further experiments on 
semiconductor
multilayer samples \cite{brookssemi}, and on the bulk quantum Hall effect in both
organic conductors \cite{organic} and an inorganic quasi-two
dimensional conductor \cite{inorganic}.

In this paper we extend the theoretical treatment of mesoscopic chiral 
conductors in two respects. First, we develop some aspects 
of the theory of conductance
fluctuations which we hope will be useful for the interpretation of 
experiments. Second, we study electron-electron interactions in the 
dirty chiral metal.

The existing understanding of conductance fluctuations in the chiral 
metal is based on calculations \cite{mathur,yu,grs2}
for the variance of sample-to-sample conductance fluctuations in systems
that are fully phase-coherent. From the results of those calculations, 
estimates have been made \cite{cho} for the variance of conductance 
fluctuations measured as a function of magnetic field in samples 
that are only partially phase-coherent, and also for the range in 
magnetic field of conductance correlations. These estimates determine 
the relevant scales but do not yield numerical coefficients or 
functional forms. Moreover, they depend on a continuum treatment 
of the system of coupled edge states, and apply if the inelastic 
scattering rate, $\tau_{in}^{-1}$ is smaller than the inter-edge 
tunneling rate, $\tau_{\perp}^{-1}$, while existing experiments 
appear to be in the opposite regime \cite{druist2,gwinntalk}. 
We supplement this
past work by calculating in full the conductance correlation 
function $\langle \delta g(B) \delta g(B+\delta B)\rangle$, 
both for $\tau_{in}^{-1}\ll \tau_{\perp}^{-1}$ and for 
$\tau_{in}^{-1}\gg \tau_{\perp}^{-1}$.

Separately, we consider conductance as a function of Fermi energy. 
For conventional disordered conductors, changes in Fermi energy 
provide a way to sample the conductance distribution. In contrast, 
we find that the conductance of a partially-coherent chiral metal 
sample, though dependent on the disorder realisation, is unchanged 
by small variations in Fermi energy. One consequence is that thermal 
smearing of the electron distribution function does not decrease the 
amplitude of conductance fluctuations in the chiral metal.

Turning to electron-electron interactions, we establish the form of the 
dynamically screened interaction and examine some of its consequences. 
Most importantly, we calculate the inelastic scattering rate due to 
electron-electron interactions, obtaining
$\tau_{in}^{-1} \propto T^{3/2}$ 
for $\tau_{in} \gg \tau_{\perp}$. 
We also examine other interaction effects which are known to be of 
interest in conventional disordered conductors\cite{altshuler}. 
Interactions in 
diffusive metals are responsible for a zero-bias anomaly 
in the tunneling density of states
\cite{aa-zba}, but -- as noted previously 
\cite{balents}, 
and as we confirm -- this is absent from the chiral metal. Similarly, 
the leading interaction contribution to the mean conductance has a 
singular temperature dependence in conventional disordered conductors
\cite{altshuler-ds1,altshuler-ds2,altshuler2}, 
but we find that it cancels in the chiral metal.

The remaining sections of this paper are organised as follows. 
We review briefly past work and collect our most important new results 
in Sec. II.
We describe calculations on conductance fluctuations in Sec. III,
and treat electron-electron interactions in Sec. IV. We summarise and discuss recent experiments in Sec. V.
\section{Background and main results}

It is useful first to introduce some notation. Consider a layered sample 
which exhibits the bulk quantum Hall effect, with one Landau level below 
the Fermi energy. A chiral metal is formed at the sample surface, from 
hybridisation of one edge state per layer. We denote the interlayer 
spacing by $a$, the chiral velocity by $v$ and the interlayer diffusion 
constant by $D$. The electron density of states per unit area is $n=1/hva$ 
and the surface conductivity in the transverse direction, as given by the 
Einstein relation, is $\sigma=D/va$, in units of $e^2/h$, which we use 
throughout this paper.

Phase coherent mesoscopic conductors are known to have three distinct 
mesoscopic regimes, according to relative sample dimensions \cite{grs2}. 
For definiteness, suppose that the sample is a cylinder of height $L$ 
and circumference $C$, with its axis perpendicular to the layers. Take 
the number of layers to be $N$, so that $L=Na$. Two intrinsic length 
scales arise, with which $L$ should be compared. The first of these 
involves the mean conductance of the sample, which if Ohm's law 
applies is $\langle g \rangle =
C\sigma/L$: the quasi one-dimensional localisation length, $\xi=2C\sigma$, 
is the value of $L$ at which $\langle g \rangle$ is of order the quantum 
unit of conductance. Samples with $L \gg \xi$ have states localised in 
the vertical direction and are quasi one-dimensional insulators.
Samples with $L \ll \xi$ are metallic and have Ohmic dependence 
of $\langle g \rangle$ on sample dimensions. A second length scale 
influences the amplitude of conductance fluctuations. It is the 
distance $L_1$ that an electron diffuses in the transverse direction 
during the time taken to circumnavigate the sample: $L_1=(DC/v)^{1/2}\equiv(a\sigma C)^{1/2}$. 
Samples with $L_1 \ll L \ll \xi$ are quasi one-dimensional metals without 
time-reversal symmetry and, in common with other examples of such systems, 
have a variance for the conductance of $\langle (\delta g)^2 \rangle =1/15$. 
Samples with $L \ll L_1$ are two-dimensional chiral metals. If contacts 
are attached to these samples for conductance measurements, electrons 
will escape by diffusion to the contacts before completing a circuit 
of the sample in the chiral direction. As a result, multiple scattering 
of an electron from a given impurity is completely supressed. Such 
samples present a number of essentially independent parallel strips 
for conduction in the transverse direction, each having a width in 
the chiral direction given by the distance an electron propagates before 
reaching a contact. This width is $L^2v/D \equiv C(L/L_1)^2$. The 
conductance of the two-dimensional chiral metal may be thought 
of \cite{mathur} as a sum of contributions from each such strip, 
and its variance has the value \cite{mathur,grs2}
$\langle \delta g^2 \rangle = L_1^2/3L^2$.

Our concern in the following is with samples that are not completely 
phase coherent. They have two different metallic regimes, analogous 
to those described above for phase coherent conductors. We restrict 
our attention to one of these, the incoherent two-dimensional metal, 
in which $\tau_{in} < C/v$ so that dephasing occurs before 
circumnavigation: this is the simpler regime theoretically 
and seems the one relevant for existing experiments. Cho, Balents 
and Fisher \cite{cho} have estimated the size of conductance 
fluctuations in the incoherent metal by viewing each phase-coherent 
region as a classical resistor and the whole sample as a random 
resistor network. We summarise their argument here for convenience. 
Let the number of regions into which the sample is divided be $n_x$ 
in the chiral direction, and $n_z$ in the transverse direction. 
Let the conductance of a single region have mean value $g_{\rm patch}$, 
with fluctuations of magnitude  $\delta g_{\rm patch}$; 
provided $\delta g_{\rm patch} \ll g_{\rm patch}$, the 
corresponding resistances are $R_{\rm patch}=1/g_{\rm patch}$ and 
$\delta R_{\rm patch}=\delta g_{\rm patch}/g^2_{\rm patch}$. 
A strip of $n_z$ such regions forms a single conduction path 
in the transverse direction, having mean resistance 
$R_{\rm strip}=n_z R_{\rm patch}$ with fluctuations 
$\delta R^2_{\rm strip} = n_z \delta R^2_{\rm patch}$; 
the corresponding conductances are $g_{\rm strip}=1/R_{\rm strip}$ 
and $\delta g_{\rm strip} = \delta R_{\rm strip}/R^2_{\rm strip}$. 
Since the sample consists of $n_x$ such paths in parallel, 
its conductance has mean value 
$\langle g \rangle = n_x g_{\rm strip} = (n_x/n_z) g_{\rm patch}$ 
and fluctuations  $\langle \delta g^2\rangle = n_x \delta g^2_{\rm strip} = 
(n_x/n_z^3) \delta g^2_{\rm patch}$. The implications 
of this final result depend on the size of a patch 
in the transverse direction, compared to the layer spacing, $a$.

If $\tau_{in}^{-1} \ll \tau_{\perp}^{-1}$ so that the phase coherence 
length in the transverse direction is much larger than the layer 
spacing, which is the case considered previously \cite{cho}, then 
one expects a continuum treatment of the system to be adequate. 
In this case, $n_z^2=L^2/D\tau_{in}$ and $\delta g^2_{\rm patch} \sim 1$; 
in addition, $n_x=C/v\tau_{in}$. Combining these expressions and 
introducing the inelastic scattering length in the chiral 
direction, $l_{in}=v\tau_{in}$, one arrives at the 
conclusion (Eq.\,(4.1) of Ref.\,\onlinecite{cho}) that
\begin{equation}
\langle \delta g^2 \rangle \sim 
\left(\frac{l_{in}}{C}\right)^{1/2}\cdot
\left(\frac{\langle g \rangle}{N}\right)^{3/2}\,.
\label{fluc1}
\end{equation} 
Experimental studies use a magnetic field with a component normal 
to the sample surface to generate conductance fluctuations in the 
chiral metal \cite{druist,druist2}; the field scale of conductance correlations 
is set by the flux quantum, $\Phi_0$,
and the area of a coherent region, being (Eq.\,(4.4) of Ref.\,\onlinecite{cho})
\begin{equation}
\delta B_0 \sim \Phi_0/(l_{in}^{3/2}a^{1/2} \sigma)\,.
\label{fluc2}
\end{equation}
The results we obtain in Sec.\,III substitute for Eqns.\,(\ref{fluc1}) 
and (\ref{fluc2}) the correlation function
\begin{eqnarray}
\nonumber
\langle \delta g(B_{\perp}) \delta g(B_{\perp}+\delta B) \rangle = 
\left(\frac{\langle g \rangle}{N}\right)^{3/2}\!\!\!\cdot 
\left(\frac{l_{in}}{C}\right)^{1/2}\!\!\!\cdot f(y) \\
\label{fluc-2d}
\end{eqnarray}
where the function $f(y)$
is 
\begin{equation}
f(y)=\pi^{-1/2} \int_{-\infty}^{\infty} d{x} e^{-({x}^2 + y {x}^6)}
\end{equation}
with the scaling variable 
\begin{equation}
y=\frac{\pi^2}{12} \cdot \sigma \cdot a l_{in}^3 \cdot 
\left(\frac{\delta B}{\Phi_0}\right)^2\,.
\end{equation}

The opposite limit of weakly coupled edges, $\tau_{in}^{-1} 
\gg \tau_{\perp}^{-1}$, has not previously been examined 
theoretically but seems to be important 
experimentally \cite{druist2,gwinntalk}. 
Adapting the argument outlined above, we take, as elements of a 
classical resistor network, regions again of length $l_{in}$ in 
the chiral direction but now of width $a$ in the transverse direction. 
Then $n_x=C/l_{in}$ and $n_z=N$. The  mean conductance of one such 
patch is small in this limit. We 
expect fluctuations to be of the same order, and take 
$\delta g^2_{\rm patch} \sim g^2_{\rm patch} =(\langle g \rangle n_z/n_x)^2$. 
In this way we obtain in place of Eq.\,(\ref{fluc1})
\begin{equation}
\langle \delta g^2 \rangle \sim \frac{\langle g 
\rangle^2}{N}\cdot\frac{l_{in}}{C}\,.
\label{fluc3}
\end{equation}
The field scale of conductance correlations is again set by the 
flux quantum and the area of a coherent region, but Eq.\,\ref{fluc2} 
is replaced in this limit by
\begin{equation}
\delta B_0 \sim \Phi_0/(l_{in}a)\,.
\label{fluc4}
\end{equation}
We calculate in Sec.\,III the conductance correlation function for 
weakly coupled edges, finding
in agreement with these estimates
\begin{equation}
\langle \delta g(B_{\perp}) \delta g(B_{\perp}+\delta B) \rangle =
\frac{{2\langle g \rangle}^2}{NC} \frac{l_{in}}{1 + z^2 }
\label{fluc-1d}
\end{equation}
where $z = 2\pi\delta B l_{in} a / \Phi_0$. 

The inelastic scattering length, $l_{in}$, appears as a phenomenological 
parameter in the expressions given above for the correlation function 
of conductance fluctuations. The contribution to inelastic scattering 
from electron-electron interactions is known in non-chiral dirty metals 
to have a characteristic temperature dependence, reflecting the form of 
the dynamically screened Coulomb interaction 
in a diffusive system \cite{altshuler,altshuler-rate,ramakrishnan,blanter}, 
and it is interesting to study inelastic scattering microscopically for 
the chiral metal. We do this in Sec.\,IV, examining dynamical screening 
and using the results to calculate the 
inelastic scattering rate, $\tau_{in}^{-1}$. 
We obtain in the regime $\tau_{in}^{-1} \ll \tau_{\perp}^{-1}$
\begin{equation}
\tau_{in}^{-1} =  c \frac{a}{D^{1/2}} 
\left(\frac{k_{\rm B}T}{\hbar}\right)^{3/2}\,,
\label{t_in}
\end{equation}
where $c\approx 1.5$ is a dimensionless coefficient.
A simple interpretation of this result can be given,
following similar arguments established in the theory of
diffusive metals \cite{altshuler}. As a starting point, note that (in both chiral and non-chiral diffusive metals) the dynamically 
screened interaction strength at long wavelengths is 
independent of the bare interaction strength. In consequence, the 
inelastic scattering rate depends only on properties of the non-interacting 
system and temperature: $\hbar \tau_{in}^{-1}$ is of order the 
single-particle energy level spacing in a system with dimensions 
determined by the distance an electron moves in the time $\hbar/k_{\rm B}T$. 
These dimensions are $L_{x}(T)\sim \hbar v/k_{\rm B}T$ and $L_{z}(T) 
\sim (\hbar D/k_{\rm B}T)^{1/2}$ in the chiral and transverse directions 
respectively, and the estimate $\tau_{in}^{-1}\sim 
(\hbar n L_{x}(T) L_{z}(T))^{-1})$ is consistent with Eq.\,(\ref{t_in}). 
An equivalent argument applied to non-chiral, diffusive conductors in 
$d$-dimensions yields $\tau_{in}^{-1}\sim T^{d/2}$, which is corrected 
in two dimensions to $\tau_{in}^{-1}\sim T|\log(T)|$ by detailed 
calculations \cite{altshuler,blanter}. The factor of $\log(T)$ in this last expression 
reflects divergent contributions to scattering in the diffusive 
two-dimensional metal, from processes with small energy 
transfer. The same scattering processes are responsible for a 
difference, in diffusive two-dimensional metals, between the 
temperature dependence of the inelastic (or out-scattering) rate, 
which acts as a cut-off for conductance fluctuations, and the 
dephasing rate, which acts as a cut-off for the weak 
localisation effects that result in negative magnetoresistance. 
In the chiral metal, as for conventional, diffusive metals in more than two 
dimensions, small energy transfer processes are not dominant and 
there is a single relaxation rate.

Inelastic scattering in weakly coupled edge states has been discussed 
in Ref.\,\onlinecite{balents}, where it is noted that
one expects $\tau_{in}^{-1} \propto T$ from perturbation 
theory in the interaction stength,
applied to a single edge.

\section{Conductance Fluctuations}

\subsection{Model}
We study conductance fluctuations in the chiral metal using a single-particle
description of the system, supplemented by a relaxation rate to represent
inelastic scattering. We treat a sample with layers lying in the $x-y$
plane and consider a surface in the $x-z$ plane. Electrons on the
surface then have a continuous coordinate, $x$, in the chiral direction
parallel to the edges of layers, and an integer coordinate, $n$, labeling
layers in the transverse ($z$) direction, which
we combine as ${\bf r}=(x,n)$. We take the 
interlayer tunneling energy to be $t$ and represent a magnetic field 
component, $B_{\perp}$, normal to the surface
using the vector potential ${\bf A}=B_{\perp}an\hat{\bf x}$. 
The Hamiltonian $H$ acts on a wavefunction
$\psi_n(x)$ according to \cite{balents,johnsondhi}
\begin{eqnarray}
\nonumber
(H\psi)_{n}(x) =  v (-i\hbar \partial_{x} + e B_{\perp}an)
 \psi_{n}(x) \\
-t[\psi_{n+1}(x)+\psi_{n-1}(x)] 
+ V_{n}(x) \psi_{n}(x)\,,
\label{schro} \\
\end{eqnarray}
where $V_{n}(x)$ is a random potential arising from impurities and 
surface roughness. We choose $V_{n}(x)$ to be Gaussian
distributed with short-range correlations, so that 
$\langle V_{n}(x)\rangle = 0$ and
$\langle V_{n}(x)V_{m}(x')\rangle = \Delta \delta_{nm} \delta(x-x')$.

Denoting the Green's function for Eq.\,(\ref{schro}) by 
$g\equiv(z-H)^{-1}$, we require its disorder average
\begin{equation}
G(z;{\bf r}_1,{\bf r}_2)\equiv \langle g(z;{\bf r}_1,{\bf r}_2) \rangle\,,
\end{equation}
the diffusion propagator
\begin{equation}
K(\omega;{\bf r}_1,{\bf r}_2)\equiv
\langle g(\omega +i0;{\bf r}_1,{\bf r}_2) g(-i0;{\bf r}_2,{\bf r}_1)\rangle\,,
\label{diff-prop}
\end{equation}
and its Fourier tranform
\begin{equation}
K(\omega,{\bf k}) \equiv \int_{-\infty}^{\infty}\!\! dx \,
\sum_n e^{i (x k_x + a n k_y)}  K(\omega;{\bf 0},{\bf r})\,.
\end{equation}
These quantities are known from calculations \cite{balents,johnsondhi}
using the usual expansion in powers of
the Green's function for the disorder-free system. 
Without disorder or a normal magnetic field component,
eigenstates are plane waves and the electron dispersion relation is
$\epsilon({\bf k})= \hbar v k_{x}- 2 t cos(k_{z}a)$. 
This dispersion along
with the chiral motion leads to an open Fermi surface on
which all electrons have the same $x$-component of velocity.
In the presence of a normal magnetic field component of
dimensionless strength $b \equiv eB_{\perp}v a/t $, 
the eigenfunctions ${\psi_n}(x)$ in the pure system are
\begin{equation}
{\psi_n}(x) = \frac{1}{\sqrt{2\pi}} e^{ik x} \phi_{\alpha}(n)\,,
\end{equation}
where $\alpha$ is an integer and 
$\phi_{\alpha}(n)=J_{n-\alpha}(2/b)$, the Bessel function
of order $n-\alpha$. The associated energy eigenvalues are
$\hbar v k + \epsilon_{\alpha}$ with $\epsilon_{\alpha}=\alpha b t$.

The treatment of disorder is simple in some respects because chiral motion
prevents multiple scattering of an electron from any particular impurity. 
As a result, the average one-particle Green's function is given exactly
by the Born approximation\cite{balents,johnsondhi}. 
In this way, one arrives at \cite{johnsondhi}
\begin{eqnarray}
\nonumber
G^R(E;{\bf r}_1,{\bf r}_2)&&=\\
\frac{1}{2\pi}\int_{-\infty}^{\infty}\!\!dk
\sum_{\alpha}&&\frac{e^{ik(x_2-x_1)}\phi_{\alpha}(n_1)\phi_{\alpha}(n_2)}
{E+i\Delta/(2\hbar v)-(\hbar v k+\epsilon_{\alpha})}\,,
\end{eqnarray}
where we use the standard notation, $G^R(E)\equiv G(E+i0)$ 
and $G^A(E)\equiv G(E-i0)$.
From the value of the self-energy, the elastic scattering time is
$\tau_{el}=\hbar^2v/\Delta$, and the elastic scattering length is 
$l_{el}=(\hbar v)^2/\Delta$.
In the absence of a normal magnetic field component,
$G^R(E;{\bf r}_1,{\bf r}_2)$ is translationally invariant and
has the Fourier transform
\begin{equation}
G^R(E;{\bf k})=
[E+i\Delta/(2\hbar v)-(\hbar v k_x-2t\cos(k_za)]^{-1}\,.
\end{equation}
For non-zero $B_{\perp}$, $G^R(E;{\bf r}_1,{\bf r}_2)$ acquires
a phase under translations, which is central to 
calculations of conductance 
fluctuations with magnetic field:
\begin{equation}
G^R(E;{\bf 0},{\bf r})=\exp(i\beta an'x)
G^R(E;{\bf r}',{\bf r}+{\bf r}')\,,
\end{equation}
where $\beta=2\pi B_{\perp}/\Phi_0$, 
${\bf r}=(x,n)$ and ${\bf r}'=(x',n')$.

The diffusion propagator is similarly given by a
a sum of ladder diagrams, and has a form that reflects the combination
of chiral motion in $x$ and diffusion in $z$.
At small wavevectors it is
\begin{equation}
K(\omega,{\bf k})= (\hbar v)^{-1}[\hbar D {k_z}^2-i(\omega+\hbar v k_x)]^{-1}\,.
\label{diffuson}
\end{equation}
The diffusion constant has the value $D=2(at)^2v/\Delta$ 
in the absence of a normal
magnetic field component, and has the field dependence \cite{johnsondhi}
$D(B_{\perp})=D(0)/[1+(B_{\perp}/B_0)^2]$, where the field scale
for magnetoresistance is $B_0=\Phi_0/al_{el}$.

Because of the simplifications in the treatment of disorder
that follow from chiral motion, the results above hold both at
weak disorder ($t \gg \Delta/(\hbar v)$) and at strong disorder
($t \ll \Delta/(\hbar v)$). 
In a semiclassical description, electron motion follows a random walk in the 
transverse direction. At weak disorder, steps of the
walk have speed $at/\hbar$ and duration $\tau_{el}$;
the rate for inter-edge tunneling is $\tau_{\perp}^{-1}=t/\hbar$. 
At strong disorder, steps are of length $a$ and duration $\tau_{\perp}$;
from the value of $D$, we deduce that the rate for inter-edge tunneling is
then 
$\tau_{\perp}^{-1}=(t/\hbar)^2 \tau_{el}$.
  
\subsection{Conductivity}
The component of the current density operator in the transverse
direction is
\begin{equation}
j_{z} ({\bf r}) = 
\frac{eat}{i \hbar} [\psi_n^{*}(x) \psi_{n+1}(x) - 
\psi_{n+1}^{*}(x) \psi_n(x)]\,.
\end{equation} 
The Kubo formula for the transverse
conductivity in a system of area $A$ is 
\begin{equation}
\sigma = \frac{ \hbar}{2 \pi A} Tr [ j_z g^R j_z g^A ]\,.
\end{equation}
Evaluating the disorder average of this expression, one finds 
\begin{equation}
\sigma = \frac{e^2}{h} \frac{D}{v a}  
\end{equation}
as expected from the Einstein relation.

\subsection{Conductance fluctuations in  magnetic field}
Conductance fluctuations have been studied 
previously \cite{mathur,yu,grs2} 
for the two-dimensional
chiral metal in the absence of inelastic scattering,
in most detail by Gruzberg, Read and Sachdev \cite{grs2},
who discussed the variance and 
its dependence on sample geometry.
Here we present calculations which include inelastic scattering, obtaining
the correlation function
\begin{equation}
F(\delta B) = \langle \delta g(B) \delta g(B + \delta B) \rangle\,.
\end{equation}
\begin{minipage}[b]{.99\linewidth}
\begin{figure}
\begin{center}
\epsfig{figure=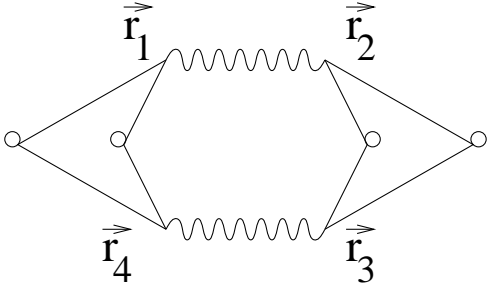,height=3.8cm}
\caption{The Feynman diagram which is relevant for the conductance 
fluctuations:
current operators are represented by circles,
one-particle Green's functions by straight lines, and
diffusion propagators by wavy lines.} 
\label{fig:diagram}
\end{center}
\end{figure}
\end{minipage}
As a result of chiral motion, and provided that electrons do not
circumnavigate the sample phase coherently ($C/v > \tau_{in}$ or 
$C/v > L^2/D$), the only one-loop diagram contributing to $F(\delta B)$
is the one shown in Fig.~\ref{fig:diagram}. It is most
conveniently evaluated in real space, and represents the expression
\begin{eqnarray}
\nonumber
F(\delta B)=\int \sum_{n_i}
K_{\delta B}({\bf r}_1,{\bf r}_2)  K^*_{\delta B}({\bf r}_4,{\bf r}_3) \\
J({\bf r}_1,{\bf r}_4)
J({\bf r}_2,{\bf r}_3) \,d\{{x}_i\}\,.
\label{fluc}
\end{eqnarray}
Here, $K_{\delta B}({\bf r}_1,{\bf r}_2)$ stands for the zero-frequency
diffusion propagator, generalised to the situation of interest, in
which the two Green's functions entering it are evaluated for
different normal magnetic field strengths,
\begin{equation}
K_{\delta B}({\bf r},{\bf r}')=\langle g_B^R({\bf r},{\bf r}')
g_{B+\delta B}^A({\bf r}',{\bf r})\rangle\,,
\end{equation}
and in which
inelastic scattering is included.
The other factors arise from a combination
of single-particle Green's functions and current operators,
and are short range: 
$J({\bf r},{\bf r}')=(\hbar\Delta^2/2\pi L^2)|C({\bf r},{\bf r}')|^2$, with
\begin{equation}
C({\bf r},{\bf r}')=\int \sum_{n_1}
G^A({\bf r},{\bf r}_1)j_z({\bf r}_1)G^R({\bf r}_1,{\bf r}') dx_1\,.
\end{equation}
Our approach to evaluating $F(\delta B)$ is different in the two regimes,
$\tau_{in}^{-1}\ll\tau_{\perp}$ and $\tau_{in}^{-1}\gg\tau_{\perp}$,
which we consider separately.

\subsubsection{Strongly coupled edges: {$\tau_{\perp}^{-1}\gg \tau_{in}^{-1}$ }}
In this regime, the discreteness of the system in the transverse
direction may be neglected. It is sufficient to approximate the short-range
terms in Eq.\,(\ref{fluc}) by  
\begin{equation}
J({\bf r},{\bf r}') = J_0 \delta({\bf r}-{\bf r}')\,,
\label{J1}
\end{equation}
where
\begin{equation}
J_0 = \frac{e^2}{h}\frac{4 (at)^2}{\Delta L^2}(\hbar v)^2\,.
\end{equation} 
The diffusion propagator,
generalised to include the magnetic field difference
$\delta B$ and inelastic scattering, is most easily calculated
within a continuum treatment of the transverse direction by solving,
for $x>0$, the differential equation
\begin{equation}
v\partial_x K_{\delta B}({\bf 0},{\bf r})=
[D\partial_z^2
+iv\delta\beta z - \tau_{in}^{-1}]K_{\delta B}({\bf 0},{\bf r})
\label{diff}
\end{equation}
with the initial condition $K_{\delta B}({\bf 0};x,z)=(\hbar v)^{-2}\delta(z)$ at $x=0$.
Here, the inelastic scattering rate, $\tau_{in}^{-1}$, has been introduced,

and the magnetic field difference enters through 
$\delta\beta\equiv(2\pi \delta B/\Phi_0)$.
To describe a sample connected to contacts, absorbing boundary
conditions ($\partial_zK_{\delta B}({\bf 0},{\bf r})=0$)
should be imposed at
$z=0$ and $z=L$. However, provided the size of a phase-coherent
region is much smaller than the sample size, so that 
$D\tau_{in}\ll L^2$, it is sufficient to use the solution
of Eq.\,(\ref{diff}) for a system of infinite extent in the $z$-direction
when evaluating the right-hand side of Eq.\,(\ref{fluc}).
This solution is
\begin{equation}
K_{\delta B}({\bf 0},{\bf r})=
(\hbar v)^{-2}\left(\frac{v}{4 \pi D x}\right)^{1/2} \exp(-S)
\label{K1}
\end{equation}
where
\begin{equation}
S=\frac{vz^2}{4Dx} +
\frac{x}{l_{in}}-\frac{i\delta \beta xz}{2} 
 + \frac{D(\delta \beta)^2x^3}{12v}\,. 
\end{equation}
Combining Eqns.\,(\ref{fluc}),\,(\ref{J1}), and (\ref{K1}), 
we obtain 
\begin{equation}
F(\delta B)=J_0^2LC\int_0^{\infty}dx \int_{-\infty}^{\infty}dz
|K_{\delta B}({\bf 0},{\bf r})|^2
\end{equation}
and so find the correlation function for 
conductance fluctuations given in Eq.(\ref{fluc-2d}) 

The form of the scaling function $f(y)$ is shown in Fig.~\ref{fig:result}.
It is normalised so that $f(0)=1$, and decreases as 
$f \propto y^{-1/6}$ for $y>>1$, so that the conductance correlation
function decays as $F(\delta B) \propto |\delta B|^{-1/3}$
for large $|\delta B|$.
\begin{minipage}[b]{.99\linewidth}
\begin{figure}
\begin{center}
\epsfig{figure=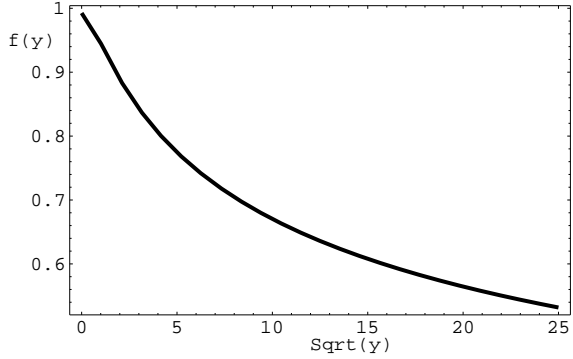,height=4.8cm}
\caption{The scaling function $f(y)$ 
\label{fig:result}}
\end{center}
\end{figure}
\end{minipage}

\subsubsection{Weakly coupled edges: {$\tau_{in}^{-1} \gg \tau_{\perp}^{-1}$}}
If adjacent edges are sufficiently weakly coupled by tunneling,
there is mesoscopic regime in which the inelastic scattering rate is small,
in the sense that $\tau_{in}^{-1} \ll \tau_{el}^{-1}$, but the interlayer
tunneling rate is even smaller. It is sufficient
in this regime to calculate the conductance correlation function
at leading order in the interlayer coupling, $t$. Doing so,
we retain $t$ in the current operators but use Green's functions
for a system of isolated edges.
The disorder-averaged single-particle Green's function in this limit

is
\begin{eqnarray}
\nonumber
G^R(E;{\bf r}_1,{\bf r}_2)= \Theta(x_2-x_1) \delta_{n_1,n_2} \times \\
(i\hbar v)^{-1}\exp(-[(2l_{in})^{-1}-i\beta an_1] [x_2 - x_1])\,,
\end{eqnarray}
where $\Theta(x)$ is the step function.
The short-range
terms associated with current operators in Eq.\,(\ref{fluc})
may then be approximated as 
\begin{equation}
J({\bf r},{\bf r}')=\frac{J_0}{2}\delta(x-x')[\delta_{n,n'+1}+\delta_{n,n'-1}]
\label{J2}\,.
\end{equation}
The diffusion propagator reduces for uncoupled edges to
a function which has a phase determined by $\delta B$
and which decays in amplitude with distance only as a result of
inelastic scattering. We take it to be
\begin{eqnarray}
\nonumber
K_{\delta B}({\bf r}_1,{\bf r}_2)= \Theta(x_2-x_1) \delta_{n_1,n_2}
\times \\
(\hbar v)^{-2}
\exp(-[l_{in}^{-1}-i\delta \beta a n_1][x_2 - x_1])\,.
\label{K2}
\end{eqnarray}
Combining Eqns.\,(\ref{fluc}),\,(\ref{J2}), and (\ref{K2}), 
we obtain 
\begin{eqnarray}
\nonumber
F(\delta B)=\frac{1}{4}J_0^2NC \times \,\,\,\,\,\,\,\,\,\,\,\,\,\,\,\,\,\,\,\,\,\,\,\,\,\,\,\,\,\,\,\,\,\,\, \\ 
\int_{0}^{\infty}dx 
[K_{\delta B}({0,n};{x,n})K_{\delta B}^*({0,n+1};{x,n+1})+{\rm c.c.}]
\end{eqnarray}
and hence arrive at the correlation function for 
conductance fluctuations given in Eq.(\ref{fluc-1d}).

\subsection{Conductance fluctuations in energy}

It is also of interest to consider the dependence of conductance
fluctuations on Fermi energy. In particular, the range of
conductance correlations in energy for a system at
zero temperature will determine the extent
to which the amplitude of fluctuations measured at finite
temperature is reduced by thermal smearing of the electron
distribution. We find that conductance fluctuations are 
perfectly correlated in energy.
To show this, we return to Eq.\,(\ref{fluc}) but replace
the magnetic field difference $\delta B$ with an energy difference 
$\delta E$. The diffusion propagator at finite energy
difference simply acquires a phase:
\begin{equation}
K_{\delta E}({\bf r},{\bf r}')=K({\bf r},{\bf r}')e^{i\delta E(x'-x)/(\hbar v)}\,.
\end{equation}
This phase cancels when 
$K_{\delta E}({\bf r}_1,{\bf r}_2)$ and 
$K^*_{\delta E}({\bf r}_4,{\bf r}_3)$ are combined, since the
factors $J({\bf r},{\bf r}')$ 
associated with the current operators are short-range,
so that ${\bf r}_1 \approx {\bf r}_4$ and
${\bf r}_2 \approx {\bf r}_3$ in Eq.\,(\ref{fluc}). As a consequence,
the correlation of conductance fluctuations is independent
of $\delta E$.
The physical reason for this behaviour is that, if $\psi_n(x)$
is an eigenfunction of Eq.\,(\ref{schro}) for some realisation
of disorder, with energy $E$, 
then $\psi'_n(x) \equiv e^{i\omega x/v } \psi_n(x)$ is 
also an eigenfunction, with energy $E'=E+\hbar \omega$,
provided $\omega C/v$ is an integer multiple of $2 \pi$,
so that periodic boundary conditions in the chiral
direction are satisfied. Hence, 
apart from the phase factor, which does not affect the conductance,  
states are perfectly correlated in energy. As a result,
the only consequence of a change in temperature is a change 
in the inelastic
scattering rate.  

\section{Effects of Interactions}

\subsection{Polarization and screening}

In this section we study several aspects of 
electron-electron interactions in the chiral metal. As a first step,
it is necessary to examine screening of the Coulomb interaction
between a pair of electrons, by other electrons in the Fermi sea.
One can expect the frequency and wavevector dependence of the 
screened interaction to be significant. In conventional 
conductors, the consequence of disorder and the resulting diffusive
motion of charge (with diffusion constant $D$)
is that screening of a potential having wavevector $q$ is suppressed \cite{altshuler}
at frequencies higher than $Dq^2$.
And in a one-dimensional chiral metal at the edge of a two-dimensional
integer quantum Hall system,
electron-electron interactions modify the dispersion relation 
for edge magneto-plasmons\cite{volkov}. 
We find in the two-dimensional chiral metal
that dynamical screening at finite wavevector
interpolates, as a function of the direction of the wavector,
between these two types of behaviour. 

Our results may be obtained either by using the random phase approximation
within a diagrammatic calculation,
or more transparently from a hydrodynamic approach, as follows.

Consider the response of the chiral metal to a space- and time-dependent
external potential. Let $\rho({\bf r},t)$ be the screening charge
density induced in the presence of an
electric field that has components $(E_x,E_z)$
within the surface. We require an equation
of motion for $\rho({\bf r},t)$, which we derive in the usual
way, by considering ${\bf J}({\bf r},t)$, the 
deviation from equilibrium of the current density,
and using the continuity equation. Treating the transverse
direction as continuous, one has
\begin{equation}
J_x({\bf r},t) = v \rho({\bf r},t)
\label{jx}
\end{equation}
and
\begin{equation}
J_z({\bf r},t) = - D \partial_z \rho({\bf r},t) + \sigma E_z\,,
\label{jz}
\end{equation}
so that the effect of the transverse component of the electric
field is simply to generate a transverse current density.
The effect of the component of the electric field in the chiral
direction is rather different: referring to the three-dimensional sample
as a whole, $E_x$ generates Hall currents 
within each layer, which 
transport charge between the bulk and the surface. The resulting
change in surface charge density may be represented by adding a term to
the continuity equation, so that it becomes
\begin{equation}
\partial_t \rho({\bf r},t) =
- {\bf \nabla} \cdot {\bf J}({\bf r},t) 
+\frac{\sigma_H}{a} E_x\,,
\label{cont}
\end{equation}
where $\sigma_H$ is the quantised Hall conductance of a layer, and
(as in preceding sections) $a$ is the layer spacing. 
It is helpful to express the conductivities in Eqns.\,(\ref{jz})
and (\ref{cont}) as $\sigma=e^2nD$ and $\sigma_H=e^2/h$, and to recall that
the density of states is $n=(hva)^{-1}$. Then, combining
Eqns.\,(\ref{jx}), (\ref{jz}) and (\ref{cont}), the charge density
evolves as
\begin{eqnarray}
\nonumber
\partial_t \rho({\bf r},t) =&&\\ 
- v \partial_x \rho({\bf r},t)
&&+  D \partial_z^2 \rho({\bf r},t)
+ e^2 n[vE_x - D \partial_z E_z].
\label{ev}
\end{eqnarray}

The electric field appearing in these equations
arises from a scalar potential $\Phi_{tot}({\bf r},t)$
which is the sum of an externally imposed potential, $\Phi_{ext}({\bf r},t)$,
and the potential $\Phi_{scr}({\bf r},t)$ due to the screening charge, 
$\rho({\bf r},t)$. These last two quantities are related in the 
standard fashion \cite{andofowlerstern} by the Poisson equation for a three-dimensional
electrostatic problem, with coordinates ${\bf r}=(x,z)$ and $y$,
in which the charge density is $\rho({\bf r},t)\delta(y)$ and
$\Phi_{scr}({\bf r},t)$ is found by evaluating the three-dimensional 
potential at $y=0$. Taking Fourier transforms, defined according to
\begin{equation}
\rho(\omega,{\bf k})\equiv \int d^2{\bf r} \int dt 
e^{i({\bf k} \cdot {\bf r} +\omega t)} \rho({\bf r},t)\,,
\end{equation}
and similarly for other quantities, one finds
\begin{equation}
\Phi_{scr}(\omega,{\bf k}) = 
U_0({\bf k})\rho(\omega,{\bf k})/e^2,\,
\label{poisson}
\end{equation}
where 
\begin{equation}
U_0({\bf k}) = \frac{e^2}{2\epsilon_r \epsilon_0|{\bf k}|}
\end{equation}
is the Fourier transform of the unscreened interaction
potential betwen a pair of electrons.
Expressing the electric field in terms of $\Phi_{tot}({\bf r},t)$
and solving the Fourier transform of Eq.\,(\ref{ev}), we find
\begin{equation}
\Phi_{tot}(\omega,{\bf k})=
\frac{\Phi_{ext}(\omega,{\bf k})}{1+
U_0({\bf k})\Pi(\omega,{\bf k})}\,,
\end{equation}
where
\begin{equation}
\Pi(\omega,{\bf k}) = n \frac{D {k_z}^2 -i v k_x }
{D {k_z}^2-i \omega - i v k_{x} }\,.
\label{pi}
\end{equation}

This result simplifies in several limits. Static screening ($\omega=0$)
is isotropic and as given by Thomas-Fermi theory,
with an inverse screening length $\kappa=e^2 n/2\epsilon_r\epsilon_0$. Dynamical 
screening of a potential with Fourier components 
only in the transverse direction ($k_x=0$) is exactly as
in a non-chiral disordered, two-dimensional metal. 
Response to a potential with Fourier components only in 
the chiral direction ($k_z=0$) is undamped, and excitations have 
the dispersion relation: $\omega = -v(k_x+\kappa)$.

In the following, we shall treat interactions in the chiral
metal using the Matsubara formalism. Within this approach,
we find for the polarisation operator\begin{equation}
\Pi(i {\omega}_n,{\bf k}) = 
n \frac{D {k_z}^2 -i v k_x {\rm sgn}({\omega}_n)}
{D {k_z}^2+|{\omega}_n| - i v k_{x} {\rm sgn}({\omega}_n)}
\end{equation}
with ${\omega}_n = 2 \pi n k_B T$,
which has, as its analytical continuation, Eq.\,(\ref{pi}).
Similarly, the screened Coulomb interaction potential is
\begin{equation}
\nonumber
U_{scr}(i\omega_n, {\bf k}) = U_{0}({\bf k})/ [1 + U_{0}({\bf k})
\Pi(i{\omega_n}, {\bf k}) ]\,.
\end{equation}

\subsection{Tunneling density of states}
The enhanced interaction between slowly relaxing density fluctuations
in a conventional, diffusive metal is responsible for
a zero-bias anomaly in the tunneling 
density of states \cite{aa-zba}.
Balents and Fisher \cite{balents} have argued that such a zero-bias anomaly
should not be expected in the chiral metal. Physically, 
charge that tunnels into the system is swept away at 
the chiral velocity; mathematically, the terms in
perturbation theory associated with this anomaly yield only
smooth contributions for the chiral metal.

In the present  subsection we examine in more detail the reasons for this.
It is sufficient for the purpose to consider a static 
interaction, $U({\bf r})$, with finite range, between
electrons separated by a distance ${\bf r}$, and to calculate
at first order in $U({\bf r})$ the disorder-averaged exchange
energy \cite{footnote}. Let $\Sigma(E)$ be the disorder-average of this
quantity for an electron
at energy $E$ interacting with a filled Fermi sea \cite{altshuler}.
The change in the density of states due to interactions is
\begin{equation}
\delta n(E) = -n \frac{d \Sigma(E)}{d E}\,.
\label{delta-n}
\end{equation}
Viewed in this way, the zero-bias anomaly in conventional,
diffusive metals arises because states lying close in energy 
have wavefunctions which are highly correlated in space,
enhancing interactions. Moverover, the wavefunction correlations
depend strongly on the energy separation of the states,
so that $\Sigma(E)$ varies rapidly with $E$ if $E$ is close to
the Fermi energy, $E_{\rm F}$.

Let $\{\Psi_{\alpha}({\bf r})\}$ and $\{E_{\alpha}\}$ be single-particle
eigenfunctions and eigenenergies in a disordered chiral metal.
The disorder-averaged exchange energy 
is
\begin{equation}
\Sigma(E)=-\frac{1}{n}\int_{-\infty}^{E_{\rm F}-E}d\omega\int d^2{\bf r}
U({\bf r}) A(\omega,{\bf r})\,,
\label{Sigma}
\end{equation}
where the two-particle spectral function, 
$A(\omega,{\bf r})$, is defined by 
\begin{eqnarray}
\nonumber
A(\omega,{\bf r})=
\langle \sum_{\alpha \beta} \delta(E_{\alpha})\delta(\omega-E_{\beta})
\Psi_{\alpha}^*({\bf 0})\Psi_{\alpha}({\bf r})
\Psi^*_{\beta}({\bf r})
\Psi_{\beta}({\bf 0})\rangle\,.
\end{eqnarray}
This spectral function can be calculated from the real part of the
diffusion propagator, Eq.\,(\ref{diff-prop}), or by considering the 
time-Fourier tranform of the density associated with a spreading wavepacket.
It has the form
\begin{eqnarray}
\nonumber
A(\omega,{\bf r})&&=\\
\frac{1}{a(\hbar v)^2}&&\left(\frac{v}{4\pi D|x|}\right)^{1/2}
\exp(-\frac{vz^2}{4D|x|}-\frac{i\omega x}{v})\,.
\label{A}
\end{eqnarray}
Combining Eqns.\,(\ref{delta-n}), (\ref{Sigma}) and (\ref{A}),
one sees that, if the potential has range $R$, then $\delta n(E)$ is independent of $E$ for 
$|E-E_{\rm F}|<\hbar v/R$.

Inspecting Eq.\,(\ref{A}), it is clear that eigenfunctions 
in the chiral metal do in fact have strong correlations: $A(\omega,{\bf r})$ 
with ${\bf r}= (x,z)$ diverges as $|x| \to 0$ for $z=0$.
These correlations, however, depend only weakly on energy,
and so do not produce a large change in the tunneling density
of states.

\subsection{Inelastic scattering rate}
The following subsection is devoted to a calculation of the 
contribution from electron-electron interactions to the inelastic
scattering rate, $\tau_{in}^{-1}$, which acts as a cut-off
for conductance fluctuations.

\begin{figure}
\begin{center}
\epsfig{figure=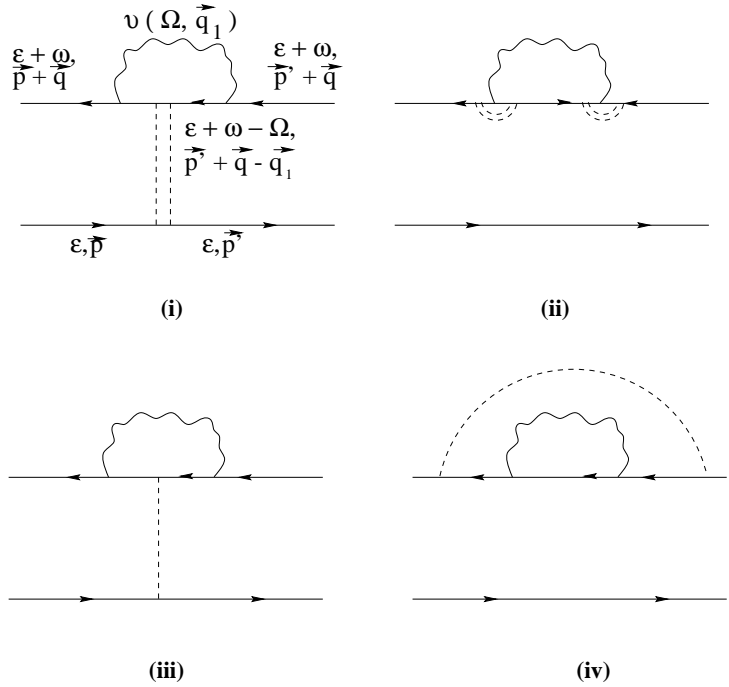,height=9cm}
\end{center}
\caption{The diagrams from which the inelastic scattering rate is calculated.}
\label{fig:InScRate2}
\end{figure}

We compute $\tau_{in}^{-1}$ by finding at leading order
the interaction contribution to the irreducible
vertex that appears in the ladder sum for the diffusion propagator
used in calculations of conductance fluctuations \cite{altshuler-rate,ramakrishnan,blanter}.
The two single-particle (advanced and retarded) Green's functions from
which this diffusion propagator is constructed are associated
with distinct measurements of the conductance. In consequence,
one should not include in the irreducible vertex diagrams
in which the interaction line connects the two single-particle 
propagators. The relevant diagrams \cite{blanter} are shown in 
Fig.~\ref{fig:InScRate2}, in which the full lines 
represent single-particle propagators, wavy lines
are the screened Coulomb interaction, single
dashed lines are impurity averages, and double dashed lines
are diffusion ladders. We evaluate these diagrams
treating the transverse direction as continuous, so that our
results are restricted to the regime in which edges are strongly
coupled: $\tau_{\perp}^{-1}\gg\tau_{in}^{-1}$. 
We set the energy 
and wavevector differences, $\omega$ and ${\bf q}$, at the vertex to zero,
express sums on Matsubara frequencies as contour
integrals, and combine terms. We expect that the dominant contribution 
to the resulting integral will be from small wavevectors and energies,
and use the asymptotic form of the integrand appropriate for
this regime, obtaining
\begin{eqnarray}\nonumber
\tau_{in}^{-1} = \frac{1}{\pi \hbar n}\int \frac{d^2 {\bf q}}{(2 \pi)^2}\int_{-\infty}^{+\infty}dx \times \\ \nonumber
[\coth(x/2k_{\rm B}T) - \tanh((x)/2k_{\rm B}T)] \times \\
\frac{x(\hbar D {q_{z}}^2)^2}{[(x+\hbar v q_{x})^2 + (\hbar D {q_{z}}^2)^2]
[(\hbar vq_{x})^2 + (\hbar D {q_{z}}^2)^2]} \,.
\end{eqnarray}
The dependence of this expression on $T$, $D$ and $v$ is made apparent by
introducing the scaled variables: $X=x/k_{\rm B}T$, $Q_x=\hbar vq_x/k_{\rm B}T$
and $Q_z=(\hbar D/k_{\rm B} T)^{1/2}q_z$. The integral can be evaluated 
numerically, yielding the expression for $\tau_{in}$ given in Eq.\,(\ref{t_in}).
The physical interpretation of this result has been
discussed in Sec.\,II.

\end{multicols}

\begin{multicols}{2}

\subsection{Interaction corrections to conductivity}
Electron-electron interactions are known in diffusive conductors
to give rise to temperature-dependent contributions to the 
conductivity that are singular in 
the low-temperature limit \cite{altshuler-ds1,altshuler-ds2,altshuler2}
The origin of these interaction corrections to the conductivity is
similar to that of the zero-bias anomaly in the tunneling density
of states. Since the chiral metal has no zero-bias anomaly,
one expects it to have no singular interaction
contribution to the conductivity. In the following subsection
we outline calculations that confirm this at leading order.
The presentation follows closely that of Altshuler, Khmel'nitskii,
Larkin, and Lee \cite{altshuler2} for the diffusive metal.

\begin{figure}
\begin{center}
\epsfig{figure=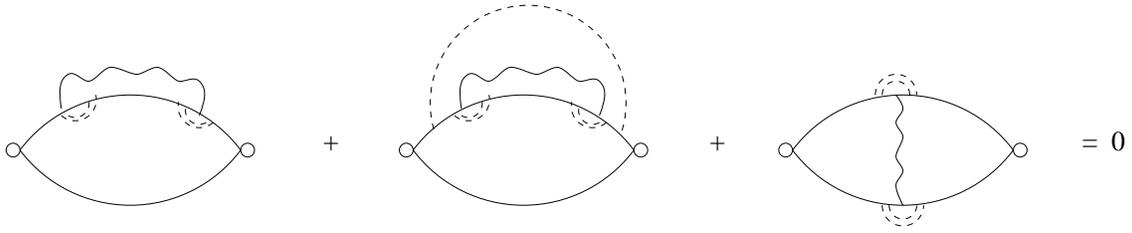,height=3cm}
\end{center}
\caption{Diagrams for the interaction correction to the
conductivity which cancel.}
\label{fig:types}
\end{figure}

\begin{figure}
\begin{center}
\epsfig{figure=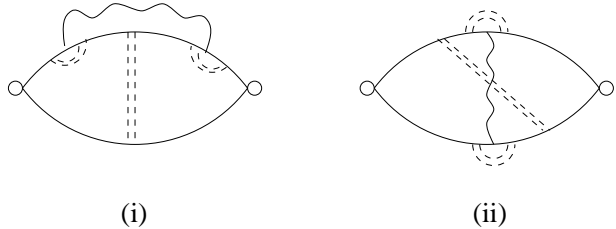,height=3.0cm}
\end{center}
\caption{The remaining leading-order diagrams 
for the interaction correction to
the conductivity.}
\label{fig:twodiag}
\end{figure}

To compute the interaction correction to the
conductivity of the chiral metal using the Matsubara
technique, it is convenient to use the Kubo formula in the form
\begin{equation}
\sigma = \lim_{\omega\rightarrow0} \frac{i\Lambda(\omega)}{\omega}\,
\end{equation}
where $\Lambda(\omega)$ is the analytic continuation in $\omega$
of 
\begin{equation}
\Lambda(i\omega_n) = \frac{k_{\rm B} T}{A}\sum_{\epsilon_m}
\langle Tr[j_z g(i[\epsilon_m+\omega_n]) j_z g(i\epsilon_m)]\rangle\,.
\end{equation}
The diagrams that contribute to $\Lambda(i\omega_n)$ at leading order
in the screened interaction strength can be classified according to the
number of diffusion propagators each one contains. Those with the
most diffusion propagators are most singular in the low-temperature
limit. They can be separated into two categories. Diagrams in the first
category are illustrated in Fig.~\ref{fig:types},
corresponding to Figs.\,5(a), 5(b) and 5(c) of Ref.\,\onlinecite{altshuler2}. 
It has been shown for a diffusive conductor in Ref.\,\onlinecite{altshuler2}
that these cancel, and we find the same to be true in the
chiral metal. Diagrams in the second category are illustrated in 
Fig.\,\ref{fig:twodiag},
corresponding to Figs.\,5(d) and 5(e) of Ref.\,\onlinecite{altshuler2}.
For the two-dimensional diffusive metal, these diagrams
make a contribution to the conductivity which is logarithmic
in temperature. For the chiral metal, their contribution 
is formally proportional to $T^{1/2}$, but with a numerical coefficient
which we find to be zero after integrating over the component in the chiral direction of the momentum transferred by the interaction.
\end{multicols}

\begin{multicols}{2}
\section{Discussion}
Our results, as set out in Sec. II, are a contribution
to the understanding of transport in the chiral metal.
In many senses, it is a particularly simple example of
a disordered conductor. Quantum interference effects are suppressed
by chiral motion, and interaction effects are less
singular than in diffusive two-dimensional conductors.
We have provided a rather detailed treatment of conductance fluctations,
which we hope will be tested experimentally. 
We have also identified a distinct mesoscopic regime,
for weakly coupled edges, in which the inter-edge tunneling rate,
$\tau_{\perp}^{-1}$, is smaller 
than the inelastic scattering rate, $\tau_{in}^{-1}$.
In this regime, transport is controlled by the properties of an
isolated, one-dimensional chiral metal.

Recent experiments \cite{druist2,gwinntalk} appear, in fact, to probe behaviour
of weakly coupled edges, though their analysis has
necessarily so far been based on theory for the opposite regime. 
According to that analysis, the inelastic scattering length
in the transverse direction, estimated from the variance of 
conductance fluctuations using Eq.\,(\ref{fluc1}), is
shorter than the inter-layer spacing, even at the lowest
temperatures investigated. Moreover, this estimate of
the inelastic scattering length
is inconsistent with the value deduced from the correlation
field for condcutance fluctuations, using Eq.\,(\ref{fluc2}),
as one might expect if the theory used is, in fact, inappropriate.
It will be interesting to find whether these discrepancies are
resolved by using the theory we have presented to analyse
the experiments. If so, such measurements provide a way
to investigate interaction effects in a one-dimensional system
of chiral fermions.

\section*{Acknowledgments}
We are particularly grateful for discussions and correspondence with
D. P. Druist and E. G. Gwinn. We also thank L. Balents, F. H. L. Essler,
I. A. Gruzberg, R. A. Smith, and A. M. Tsvelik for useful discussions.
J.B. 
acknowledges partial financial support from the European
Union through the Marie Curie Grant; the work was also supported in part by EPSRC under Grant No GR/J8327.

\end{multicols}

\begin{thebibliography}{99}
\bibitem{footnote} There is no Hartree correction to the density of states
in the chiral metal because impurity scattering does not generate
density fluctuations.
\end{thebibliography}
\end{document}